\documentclass[preprintnumbers,amsmath,amssymb,floatfix,superscriptaddress,nofootinbib]{revtex4}
\pdfoutput=1
\usepackage{amsmath,hyperref,amssymb}
\usepackage{graphicx}% Include figure files
\usepackage{dcolumn}% Align table columns on decimal point
\usepackage{bm}% bold math
\usepackage{verbatim}
\usepackage{tabularx}
\usepackage{slashed}
\usepackage{cancel}
\usepackage[boxsize=2em,aligntableaux=center]{ytableau}
\usepackage{float}

\def\be{\begin{equation}}
\def\ee{\end{equation}}
\def\bea{\begin{eqnarray}}
\def\eea{\end{eqnarray}}

\newcommand{\gev}{{\rm GeV}}

\begin{document}

\title{Relaxation from particle production}
\author{Anson Hook and Gustavo Marques-Tavares}
\affiliation{\,Stanford Institute for Theoretical Physics, Stanford University, Stanford, CA 94305, USA \vspace{2pt}}

\begin{abstract}
\begin{centering}
{\bf Abstract}\\[4pt]

We consider using particle production as a friction force by which to implement a ``Relaxion" solution to the electroweak hierarchy problem.
Using this approach, we are able to avoid superplanckian field excursions and avoid any conflict with the strong CP problem.  The relaxation mechanism can work before, during or after inflation allowing for inflationary dynamics to play an important role or to be completely decoupled.

\end{centering}

\end{abstract}

\vspace*{1cm}

\maketitle

\section{introduction}

There has recently been a renewed interest in solutions to the electroweak hierarchy problem which do not have new particles rendering the Higgs mass UV insensitive. The canonical example of a theory of this kind has been the multiverse, where the existence of many vacua with different Higgs masses results in a vacuum with the correct Higgs mass.  Anthropics is then used to explain why we are in that particular vacuum.  More recently, two new approaches have been proposed, the ``Relaxion"~\cite{Graham:2015cka} and $N$Naturalness~\cite{Arkani-Hamed:2016rle}.  In $N$Naturalness, instead of many different vacua, many different copies of the SM are used to give a small Higgs mass.  Reheating proceeds in such a way that only the copy with the lightest Higgs mass is reheated.  In the relaxion approach there are many vacua so that some of them have the correct Higgs mass.  Given reasonable initial conditions, the cosmological evolution of the relaxion field results in a long period of time with the correct Higgs mass.

In this letter, we take the relaxion approach to solving the hierarchy problem.  The relaxion is a novel approach where the evolution of a field $\phi$ scans the Higgs mass parameter.  Feedback from the Higgs vev crossing 100 GeV causes $\phi$ to stop in a minimum where the Higgs vev is near 100 GeV.  A crucial aspect of the relaxion approach is the presence of a frictional force, which removes all of the kinetic energy gained by the relaxion as it scans its potential.  This is important so that the relaxion can be affected by the small Higgs vev.
 
There are already a number of relaxion models in the literature~(see e.g.~\cite{Espinosa:2015eda,Hardy:2015laa,Patil:2015oxa,Batell:2015fma,Matsedonskyi:2015xta,Evans:2016htp} and references therein), most of which rely on Hubble friction in order to dissipate the relaxion energy, allowing it to be trapped in the correct minimum\footnote{See Ref.~\cite{Hardy:2015laa} for another attempt which doesn't use Hubble friction.}. In this paper we explore a new approach to the relaxion by using particle production rather than the expansion of the universe to provide a friction force. This allows us to solve several aesthetic and maybe even theoretically troublesome aspects of the relaxion scenario.  Firstly, we are able to avoid superplanckian field excursions by the relaxion field. Superplanckian field excursions have long been theoretically suspect.  The first realization of this worry comes from Giddings and Strominger~\cite{Giddings:1987cg} who showed that a periodic scalar with period f and no potential has gravitational instantons whose action scales as $S \sim M_p/f$ so that non-perturbative effects are important if $f \gtrsim M_p$.  It remains an open question whether or not this objection applies to particles more interesting than free periodic scalars or even if these non-perturbative effects should be included in the sum over topologies.

Another troublesome aspect of the original relaxion approach is that a low scale of inflation as well as large amounts of inflation was unavoidable due to the need to transverse large field distances when slowly rolling~(in App.~\ref{App: N}, we present a simple lower bound on the number of e-foldings).  There is a theory bias against low scale inflation simply because of the model building difficulty of obtaining large amounts of inflation without fine-tuning the inflaton potential.
Large amounts of inflation can be troubling because our semi-classical description of inflation breaks down if $N \gtrsim M_p^2/H^2$~\cite{ArkaniHamed:2007ky}.  While the arguments for this are based on quantum information, in the context of slow roll inflation, one can show that having $N \gtrsim M_p^2/H^2$ always leads to eternal inflation.  This places additional constraints on relaxion models that have sometimes been neglected, leading to benchmark points which have inflaton sectors that are eternally inflating.  When using particle production over Hubble friction, we will find that in certain scenarios that inflationary dynamics can be completely decoupled from the relaxation mechanism and thereby removing any constraints on inflation.

Having espoused the virtues of using particle production, we now briefly outline our model for the simple case of the Abelian Higgs in the $M_p \rightarrow \infty$ limit.  The Lagrangian is
\bea
\mathcal{L} \supset (\Lambda^2 - \epsilon \phi ) h h^\dagger + \epsilon \phi \Lambda^2 + \Lambda_c^4 \cos \frac{\phi}{f'} + \frac{\phi}{f} F \tilde F
\eea  
where $\Lambda$ is the UV cutoff and $\Lambda_c$ is an axion like potential generated from the confinement of some other gauge group\footnote{Unlike the case of the relaxion, this potential is not linked to the Higgs vev in any way}. The parameter $\epsilon$ is a spurion that controls the breaking of the shift symmetry that protects the $\phi$ potential.
The higher dimensional coupling to the U(1) gauge field will be the source of particle production.  As will be shown in Sec.~\ref{Sec: pp}, this coupling leads to exponential particle production if 
\bea
\label{Eq: max speed}
\dot \phi \gtrsim f m_A
\eea
where $m_A$ is the gauge boson mass, due to some modes of the gauge field $A$ becoming tachyonic.

Initial conditions are taken to be $\phi_0 \sim \Lambda^2/\epsilon$, $\dot \phi_0 \ne 0$ and the Higgs vev is taken to be at its instantaneous minimum.  As long as $\dot \phi \gtrsim \Lambda_c^2$, the $\phi$ field moves along without stopping at any of the minima.  Initially, the Higgs mass is large and negative and thus the gauge field is higgsed.  Because the mass of the gauge boson is large, Eq.~\ref{Eq: max speed} is not satisfied and no particle production occurs.  Eventually the Higgs mass decreases enough such that Eq.~\ref{Eq: max speed} is satisfied and particle production occurs.  From energy conservation we know that the energy for these new particles comes from the kinetic energy of $\phi$. This energy loss gives an effective friction term for $\phi$ originating from its interaction with the gauge fields.  At this point, $\phi$ loses its momentum very efficiently and becomes trapped in the nearest minimum.  While the details of implementing this mechanism before, during and after inflation are all different (especially in how they treat the initial $\dot \phi$) the previous discussion is the general mechanism for how $\phi$ stops at the correct Higgs minimum.

In the remainder of this letter, we fill in the details of implementing relaxation with particle production.  In Sec.~\ref{Sec: pp}, we show how an axion like coupling can lead to exponential particle production.  In Sec.~\ref{Sec: SM} we discuss how to use particle production to relax the SM Higgs mass to its observed small value.  In Sec.~\ref{Sec: variants}, we discuss simple variations of the models discussed in Sec.~\ref{Sec: SM}.  In Sec.~\ref{Sec: discussion}, we discuss the results.

\section{Particle production}
\label{Sec: pp}

In this section, we discuss how the coupling
\bea
\label{Eq: axion}
\mathcal{L} \supset -\frac{\phi}{4 f} F \tilde F
\eea
can cause particle production.  The fact that this coupling leads to particle production has been observed and used many times before~\cite{Anber:2009ua,Barnaby:2010vf,Barnaby:2011vw,Durrer:2010mq}. We first study particle production at zero temperature and later introduce the effects of non zero temperature.

\begin{itemize}
\item $T = 0$
\end{itemize}

Our starting point is a field $\phi$ with the shift symmetric coupling shown in Eq.~\ref{Eq: axion}.  We first examine the equations of motion for the higgsed gauge field $A$.  Working in unitary gauge we have $\partial_\mu A^\mu = 0$. For a spatially homogenous $\phi$, the coupling in Eq.~\ref{Eq: axion} only affects the transverse modes of $A$, which we write in terms of the two circular polarizations $A_\pm$.  The equations of motion for the two transverse polarizations are
\bea
\label{Eq: diffeq}
\ddot A_\pm + (k^2 + m_A^2 \pm \frac{k \dot \phi}{f} ) A_\pm = 0 
\eea
In the limit where particle production does not yet backreact on $\phi$ and one can treat $\dot \phi$ as a constant, one finds the simple expression for $A_\pm$
\bea
A_\pm (k) \propto e^{i w_\pm t} \qquad w^2_\pm = k^2 + m_A^2 \pm \frac{k \dot \phi}{f}
\eea
From this expression it is clear that there is a tachyonic growing mode for some value of k if
\bea
w_\pm^2 = k^2 + m_A^2 \pm \frac{k \dot \phi}{f}  < 0 \qquad \Rightarrow \qquad | \dot \phi | \gtrsim 2 f m_A
\eea
This tachyonic mode drains the energy of the $\phi$ field exponentially quickly.  As this coupling is shift symmetric, it is clear that the energy necessary for exponential particle production comes at the expense of $\dot \phi$ as $\phi$ might not even have a potential to take energy out of.

\begin{itemize}
\item $T \ne 0$
\end{itemize}

In the course of this paper, we will need to examine the behavior of Eq.~\ref{Eq: axion} in the presence of a thermal bath.  As such, we generalize the previous discussion to the case of a large temperature $T$.  For a thermal bath, there are several important points.  The first is that if the temperature is larger than the weak scale, gauge symmetry is linearly realized rather than non-linearly realized.

Another important difference is that at finite temperature the system is in a plasma state, and thus charged particles screen electric fields and slow the tachyon growth. This can be seen by studying the dispersion relation at finite temperature
\bea
\omega^2-k^2 \mp \frac{k \dot \phi}{f} = \Pi_t (\omega,k)
\eea
At finite temperature, the gauge bosons have longitudinal and transverse modes.  The longitudinal modes do not have a tachyon so we study the transverse modes.

We are interested in the case $\omega = i \Omega$, i.e. a tachyon, and in the regime $ | \Omega | \ll |k| \ll T$. In this case we can write the dispersion relation at 1-loop as~\cite{Kraemmer:1995qe,Kapusta:2006pm}
\bea
\label{Eq: dispersion}
- \Omega^2 - k^2 \mp \frac{k \dot \phi}{f} = \frac{m_D^2}{2} \left( \frac{i \Omega}{k} \right)\left[ \frac{1}{2}\left(1+\frac{\Omega^2}{k^2}\right)\log \left( \frac{i \Omega+k}{i \Omega-k} \right) +\frac{i \Omega}{k}\right] \approx \frac{m_D^2   |\Omega | \pi}{4 k},
\eea
where $m_D \sim g T$ is the Debye mass of the plasma. 

One sees from the equation above that there are always tachyonic modes for one of the two polarizations of $A$ and that the tachyon frequency is
\bea
\label{Eq: tachyon-temperature}
\Omega \sim \frac{ \left( \frac{| \dot \phi |}{f} \right)^3}{m_D^2},
\eea
with $k \sim \dot \phi /f$. Therefore there is an associated growing mode for the gauge field
\bea
\label{Eq: therm growth}
A \sim e^{\Omega t}
\eea

As we have only worked at the 1-loop level, one might worry that higher loops may remove the tachyon as the temperature is parametrically larger than the size of the tachyon.  To investigate this, one can Taylor expand the gauge boson self energy
\bea
\Pi_t(\omega,k) \approx \Pi_t(0,k) + \frac{\partial \Pi_t}{\partial \omega} (0,k) \omega + \cdots
\eea
In order for Eq.~\ref{Eq: dispersion} to be the leading order dispersion relation, we need $\Pi_t(0,k) = 0$.  $\Pi_t(0,k)$ is known as the magnetic mass for the gauge field, although it has nothing to do with any explicit mass term in the Lagrangian.  As argued by Linde~\cite{Linde:1980ts,Gross:1980br}, in non-abelian plasmas a magnetic mass of order $g^2 T$ is generated\footnote{See Ref.~\cite{Espinosa:1992kf} for discussion of this mass in the SM.}.  This magnetic mass stabilizes the tachyon so that non-abelian gauge fields at finite temperature will not be exponentially produced.  The case is different for abelian gauge groups where magnetic masses are known not to be generated~\cite{Fradkin}.  Thus we find that the exponential particle production at high temperatures only happens for $U(1)$ gauge fields.  

We see that there are important differences between the high temperature ($T>v$) and zero temperature cases: in the finite temperature abelian case the Higgs expectation value falls to zero and the time scale associated with the tachyon is longer due to plasma effects.  As will also be critical, at zero temperature one does the analysis in the mass basis while at high temperature, when the gauge symmetry is restored, one does the analysis in the gauge basis.

Later in the paper, we will wish to estimate how long it takes for $\dot \phi$ to fall below some critical value $\dot \phi_c$.  As can be seen from Eq.~\ref{Eq: tachyon-temperature}, the process of losing energy is IR dominated.  The smaller $\dot \phi$ is, the longer it takes to lose an $\mathcal{O}(1)$ fraction of its energy.  Thus the time it takes for $\dot \phi$ to fall below $\dot \phi_c$ is roughly
\bea
\label{Eq: time}
t \sim \frac{T^2 f^3}{\dot \phi_c^3}
\eea
This time dependance is expected purely from dimensional analysis, and is in accordance with the non-thermal case which can be studied analytically~\cite{Anber:2009ua}. A more accurate estimate including order one numbers requires a dedicated numerical investigation and is beyond the scope of this work.

\section{Relaxation with particle production}
\label{Sec: SM}

In this section, we work out the details of implementing the relaxation scenario using particle production.  As the basic mechanism behind implementing this before, during or after inflation is the same, we first give a general account of the mechanism.  We start with the Lagrangian 
\bea
\label{Eq: L}
\mathcal{L} \supset (\Lambda^2 - \epsilon \phi ) h h^\dagger + \epsilon \phi \Lambda^2 + \Lambda_c^4 \cos \frac{\phi}{f'} - \frac{\phi}{f} ( \alpha_Y B \tilde B  - \alpha_2 W \tilde W )
\eea 
The field $\phi$ is the relaxion, a periodic scalar with period $\gtrsim \Lambda^2/\epsilon$, which scans the Higgs mass. To avoid confusion, throughout this section we will refer to the Higgs boson as $h$, Hubble as $H$ and use $v$ for the final Higgs vev. The couplings $\alpha_Y$ and $\alpha_2$ are the fine structure constants for $U(1)_Y$ and $SU(2)_W$.  As they are $\mathcal{O}(1)$, they will be omitted in the remainder of the text. 

Aside from the coupling to the Higgs\footnote{%
The coupling with the Higgs can be either linear or quadratic.  Given that $\phi$ is a CP odd scalar it is more consistent with symmetries to write a quadratic interaction.  As a linear term is always generated from expanding $\phi$ around values away from zero, we will simply write down a linear term leaving it implicit that it came from a quadratic interaction.}%
, the relaxion also has a cosine potential generated from some other confining gauge group.  The large difference in periods of the two cosine potentials (with periods $f'$ and $\Lambda^2/\epsilon$) is responsible for generating many different vacua some of which have the correct Higgs mass.  One can obtain these different periodicity through any number of means~\cite{Choi:2015fiu,Kaplan:2015fuy,Fonseca:2016eoo}, however from an effective field theory point of view, we do not need to worry about such things and thus we shall not. In order to have multiple minima we require
\bea
\label{Eq: minima-exist}
\epsilon \Lambda^2 f' < \Lambda_c^4 \
\eea
otherwise the linear slope would be too steep and there would be no minima of the cosine potential.

We will take the initial conditions $\dot \phi_0 \gtrsim \Lambda_c^2$ and $\phi_0 \sim \Lambda^2/\epsilon$ such that the Higgs has a large negative mass.  We assume that the Higgs is at its minimum with a vev $\langle h \rangle$ and that the temperature of the universe is negligible.  More detailed discussions of these initial conditions and how one might obtain them dynamically under various conditions will be discussed in the following subsections.
Initially, $\phi$ has a large kinetic energy and flies over the small $\Lambda_c$ sized bumps.  Because there is already $\Lambda^4$ potential energy density in the $\phi$ field, we know that
\bea
\label{Eq: H}
H \gtrsim \frac{\Lambda^2}{M_p}
\eea
In order to scan the Higgs mass with enough precision to obtain the EW scale Higgs mass, we must also impose
\bea
\label{Eq: fine-scan}
\epsilon f' < v^2
\eea

Note that in Eq.~\ref{Eq: L}, that $\phi$ couples to $B \tilde B  - W \tilde W$.  This is critical because this combination does not contain the photon.  The photon is massless regardless of the Higgs vev and the exponential production of the photon would be insensitive to the Higgs vev and ruin the mechanism.  This coupling could naturally come out of the UV completion.  For example, a left right symmetric UV completion where $\phi$ is a P even and CP odd scalar and thus must couple to $W_R \tilde W_R - W_L \tilde W_L$, which after left right symmetry breaking gives us the operator of interest.  We give more details in App.~\ref{App: LR} and estimate how large of a coupling is allowed in Sec.~\ref{Sec: additional concerns}.

As mentioned in Sec.~\ref{Sec: pp}, our mechanism only works at finite temperature only when using abelian gauge bosons.  
At first glance, the lack of a coupling to the photon would mean that the coupling shown in Eq.~\ref{Eq: L} ceases to cause exponential particle production at finite temperature.  However as mentioned before, the tachyonic mode at finite temperature is not associated with the $SU(2)_W$ gauge bosons but instead with the $B$ gauge boson.  $\phi$ couples to the $B$ and thus exponential particle production still continues at finite temperature.  Here is it critical that the tachyonic modes at zero and finite temperatures are different.

As shown in Sec.~\ref{Sec: pp}, particle production occurs when 
\bea
\label{Eq: speed}
\dot \phi \gtrsim f m_W \sim f \langle h \rangle
\eea
Initially Eq.~\ref{Eq: speed} is not satisfied.  Once the Higgs mass becomes sufficiently small, such that $ \langle h \rangle \sim \dot \phi_s / f \sim 100$ GeV, the relaxion looses energy by exponential production of gauge bosons.  $\dot \phi_s$ is the kinetic energy right before particle production starts.  Because $\phi$ may have gained energy when rolling down its potential, $\dot \phi_s$ is not necessarily $\dot \phi_0$.  Obtaining the correct Higgs vev implies
\bea
f \sim \frac{\dot \phi_s}{v}
\eea
where $v$ is the electroweak vev in our Universe. 

The produced gauge bosons have energies comparable to $v$ and quickly reach thermal equilibrium (see Sec.~\ref{Sec: additional concerns} for more details). Thermal backreaction is critical in order for $\phi$ to lose all of its kinetic energy once particle production starts to occur.  Once particle production occurs, the Higgs obtains a finite temperature mass which is different from its vacuum value.  Due to the coupling
\bea
\mathcal{L} \supset g^2 H H^\dagger A_\mu  A^\mu
\eea
the Higgs obtains a positive contribution to its mass due to thermal and finite density effects.  Thus as particles are produced, the Higgs vev decreases.  As the Higgs vev decreases, Eq,~\ref{Eq: speed} is satisfied to a better and better degree.  Eventually when the energy density in the gauge field reaches $\rho_A \sim v^4$ (this is to be compared to the larger $\dot \phi^2$ energy density in $\phi$) the Higgs mass becomes positive.  At this point, no matter the value of $\dot \phi$, there will be a tachyon for the massless gauge boson $B_\mu$.  As a result, $\dot \phi$ approaches zero with a time dependence estimated in Sec.~\ref{Sec: pp} where
\bea
m_D \sim g' T \sim \sqrt{\dot \phi_s}
\eea
Once $\phi$ has lost its kinetic energy, it gets trapped in the cosine potential, at a value associated with a small Higgs mass.

We now discuss various constraints that this model must satisfy that limit how large the UV cutoff can be taken to be.  A lower bound on $\epsilon$ can be taken by requiring subplanckian field excursions.
\bea
\label{Eq: epsilon}
M_p > \frac{\Lambda^2}{\epsilon} 
\eea
Another constraint is that particle production needs to be able to stop the $\phi$ rolling.  $\phi$ stops rolling when its kinetic energy 
$\dot \phi$ falls below $\Lambda_c^2$.  In order for $\dot \phi$ to reach $\Lambda_c^2$, it needs to lose energy faster than it gains it through the small linear slope (the cosine contribution averages out). The energy loss is most effective if the particle production is exponential, which requires the time scale for particle production to be faster than Hubble, and thus
\bea
\frac{\Lambda_c^6}{f^3 T^2} \gtrsim H,
\eea
where we used the slowest time scale for particle production when $\dot \phi \sim \Lambda_c^2$. In this regime $\phi$ loses an order one fraction of its kinetic energy in the time scale associated with the tachyon and thus the condition for losing energy efficiently becomes
\bea
\frac{d E}{d t} < 0 \qquad \rightarrow \qquad  \Lambda_c^8 > \frac{\epsilon \Lambda^2}{v^3} \, \dot \phi_s^4 \gtrsim \frac{\Lambda^4}{M_p v^3} \, \dot \phi_s^4  \ ,
\eea
which imposes a lower bound on $\Lambda_c$. The final constraint we will discuss is that $\phi$ does not overshoot the correct Higgs mass in the time it takes for it to lose all of its kinetic energy.  In the time it takes for $\dot \phi$ to fall below $\Lambda_c^2$, the mass of the Higgs has changed
\bea
\label{Eq: pre overshoot}
\Delta m_H \sim \frac{\Delta m_H^2}{m_H} \sim \frac{\epsilon}{v} \int dt \dot \phi
\eea
The time dependent velocity of $\phi$ can be found in Eq.~\ref{Eq: time}.  Integrating Eq.~\ref{Eq: pre overshoot}, until $\dot \phi \sim \Lambda_c^2$,  we find the constraint
\bea
\label{Eq: overshoot}
\frac{\epsilon f^3 m_D^2}{v \Lambda_c^4} \lesssim v
\eea
Combining Eq.~\ref{Eq: epsilon} and Eq.~\ref{Eq: overshoot} we find that there is an upper bound on the cutoff which is
\bea
\label{Eq: bound}
\Lambda^2 \dot \phi^4_s \lesssim M_p v^5 \Lambda_c^4
\eea
This equation places the main constraint on the UV cutoff.  A conservative estimate can be imposed using $\dot \phi_s \lesssim \Lambda^2$ and $\Lambda_c \lesssim \Lambda$ giving the conservative constraint $\Lambda \lesssim (M_p v^5)^{1/6} \sim 5 \times 10^4 \ \gev$. 
The only difference between the various implementations is what value of $\dot \phi_s$ to take and what role does Hubble play in the story.

We now present several subsections where we clarify how things proceed when implementing the above mechanism before, during and after inflation. We provide an example point in parameter space for each of these scenarios as a rough guide for the typical size of the different parameters.  Additionally, we address some other concerns and show that there are no additional problems.

\subsection{Relaxation before inflation}
\label{Sec: before}

\begin{table}[h]
\centering
\begin{tabular}{|c|c|c|c|c|c|c|} 
\hline 
 & $\Lambda$ & $H$ & $\epsilon$ & $f$ & $f'$ & $\Lambda_c$ \\
\hline 
Values in GeV & $10^4$ & $10^{-10}$ & $10^{-10}$ & $10^6$ & $10^{14}$ & $10^3$ \\
\hline
\end{tabular}\caption{A sample point for relaxation before inflation, which satisfies all the inequalities (and saturates many of them).}\label{table: before}
\end{table}

We now consider the situation if relaxation occurs before inflation starts or during its first e-folding or so.
For $\dot \phi \sim \Lambda^2$, it takes a time $t \sim 1/\epsilon$ for $\phi$ to explore its entire field range.  In order for Hubble friction to be negligible, we take 
\bea
\label{Eq: EgtrH }
\epsilon > H
\eea
This means that the time scale for finding the correct Higgs minimum is much faster than a single Hubble time.  

The simplest setup for relaxation before inflation is to start with the following set of (strange) initial conditions:
\bea
\begin{aligned}
\label{Eq: unnatural-ic}
\phi_0 & \sim \Lambda^2/\epsilon ,  \\
\dot \phi_0 & \sim \Lambda^2 , \\
\langle h \rangle & \, \, \text{at its minimum} \\
T & \sim 0.
\end{aligned}
\eea
With this setup the scanning occurs as described in the previous section, and from Eq.~\ref{Eq: bound} we find $\Lambda^{10} < v^5 \Lambda_c^4 M_p$. This leads to a maximum cutoff of around $5 \times 10^4$ GeV, when $\Lambda_c \sim \Lambda$.  These initial conditions might come from a tunneling event to our vacuum.

\subsection{Relaxation during inflation}
\label{Sec: during}

\begin{table}[h]
\centering
\begin{tabular}{|c|c|c|c|c|c|c|c|} 
\hline 
 & $\Lambda$ & $H$ & $\epsilon$ &  $N_e$ & $f$ & $f'$ & $\Lambda_c$ \\
\hline 
Values in GeV & $10^5$ & $10^{-5}$ & $10^{-6}$ & $10^2$ & $3\times 10^6$ & $10^{9}$ & $1.5 \times 10^4$ \\
\hline
\end{tabular}\caption{A sample point for relaxation after inflation which satisfies all inequalities (and saturates many of them).  Rather than choosing a data point that allows for the highest UV cutoff possible, we chose a data point which has only 100 e-foldings of inflation.}\label{table: during}
\end{table}

The scanning mechanism can also work during inflation if $\phi$ is ``slow rolling'' ($\epsilon < H$). If $\phi$ were to actually slow roll it would simply get stuck in the first minimum of the cosine it encounters. However, if the amplitude of the cosine potential is small compared to the cutoff and its frequency is not too large, $\phi$ can go over the bumps and gain enough energy from the linear slope to compensate for Hubble friction. 
In fact, one can show that under certain conditions  the relaxion approaches an effective slow-roll solution at long times characterized by
\bea
\label{Eq: almost-slow-roll}
\dot \phi \sim \frac{ \epsilon \Lambda^2}{3 H} + \delta (t)
\eea
Where $\delta$ is a small oscillatory contribution.

Using Eq.~\ref{Eq: almost-slow-roll} and the equation of motion for $\phi$, one can show that there are three conditions that must be satisfied to get to this approximately slow roll solution.  These conditions are
\begin{itemize}
	\item Start with sufficiently large kinetic energy to go over the barriers%
	\footnote{One can relax this condition if there is a coincidence of scales $\Lambda_c^4 \sim \epsilon \Lambda^2 f'$, in which case a generic point in $\phi$ space will roll over the barriers even if it started at rest.}: $\dot \phi_0 \gtrsim \Lambda_c^2$.
	\item Average slow-roll velocity must be large enough to go over barriers: $ \frac{\epsilon \Lambda^2}{3 H} \gtrsim \Lambda_c^2$.
	\item Time scale to go over one period of the cosine potential must be fast compared to Hubble: $H < \sqrt{\frac{\epsilon \Lambda^2}{f'}}$.
\end{itemize}
One can show, see Eq.~\ref{Eq: minima-exist}, that the second condition is always stronger than the third. If these conditions are satisfied, the relaxion evolution approaches the almost slow roll behavior of Eq.~\ref{Eq: almost-slow-roll} quickly.

The kinetic energy of $\phi$ is always below the cutoff, since $\epsilon < H$ in this scenario. The number of e-folds required to effectively scan the Higgs mass is given by
\bea
\label{Eq: nefold}
N_e \sim \left( \frac{H}{\epsilon} \right)^2
\eea
If inflation ends before the relaxion finishes scanning the Higgs mass, the scanning process can still continue after inflation as discussed in the next section.  There is an upper bound on the number of e-foldings required to finish scanning which comes from $f \gtrsim \Lambda$ and $\dot \phi \sim f v$.  This gives us the bound that
\bea
\label{Eq: Nbound}
N_e \lesssim \frac{\Lambda^2}{v^2}
\eea

Once the relaxion reaches the quasi slow roll behavior, the scanning mechanism happens as was described in the beginning of Sec.~\ref{Sec: SM}, with $\dot \phi_s = \epsilon \Lambda^2 / 3H$. 
As in the almost instantaneous scanning of the previous section, the main constraint in this scenario comes from not overshooting the correct Higgs mass, given the large velocity of $\phi$. Using Eq.~\ref{Eq: bound} and Eq.~\ref{Eq: Nbound} we find
\bea
\label{Eq: inflation-bound}
\Lambda^4 \lesssim M_p v^3 \qquad \Lambda \lesssim 10^6 \ \gev
\eea

\subsection{Relaxation after inflation}

\begin{table}[h]
\centering
\begin{tabular}{|c|c|c|c|c|c|c|} 
\hline 
 & $\Lambda$ & $\epsilon$ & $f$ & $f'$ & $\Lambda_c$ \\
\hline 
Values in GeV & $10^4$ & $10^{-5}$ & $10^6$ & $10^{9}$ & $10^3$ \\
\hline
\end{tabular}\caption{A sample point for relaxation after inflation, which satisfies all inequalities (and saturates many of them).}\label{table: before}
\end{table}

Relaxation can also occur after inflation.  In order for relaxation to occur after inflation, we need our initial conditions to be obtainable in reasonable models of inflation.  After inflation and before reheating, one naturally has $T=0$. We focus on cases in which the Higgs mass is at least of order Hubble during inflation, either because Hubble is small compared to the cutoff as in the previous sections, or because the renormalizable $h h^\dagger R$ coupling is $\mathcal{O}(1)$. This causes the Higgs to relax to the minimum of its potential during inflation.  Thus, the only initial condition that is not satisfied is the large $\phi$ velocity. 

The large $\phi$ velocity as an initial condition was needed in order for $\phi$ to scan many vacua. As discussed in the previous section, the relaxion could exit inflation with a steady state velocity that allows for $\phi$ to go over the bumps, see Eq.~\ref{Eq: almost-slow-roll}. In this case, if inflation ends before the relaxion had time to scan the Higgs mass, one has the right initial conditions to finish the scan after inflation.

\begin{figure}[t]
\begin{center}
\includegraphics[width=0.4\textwidth]{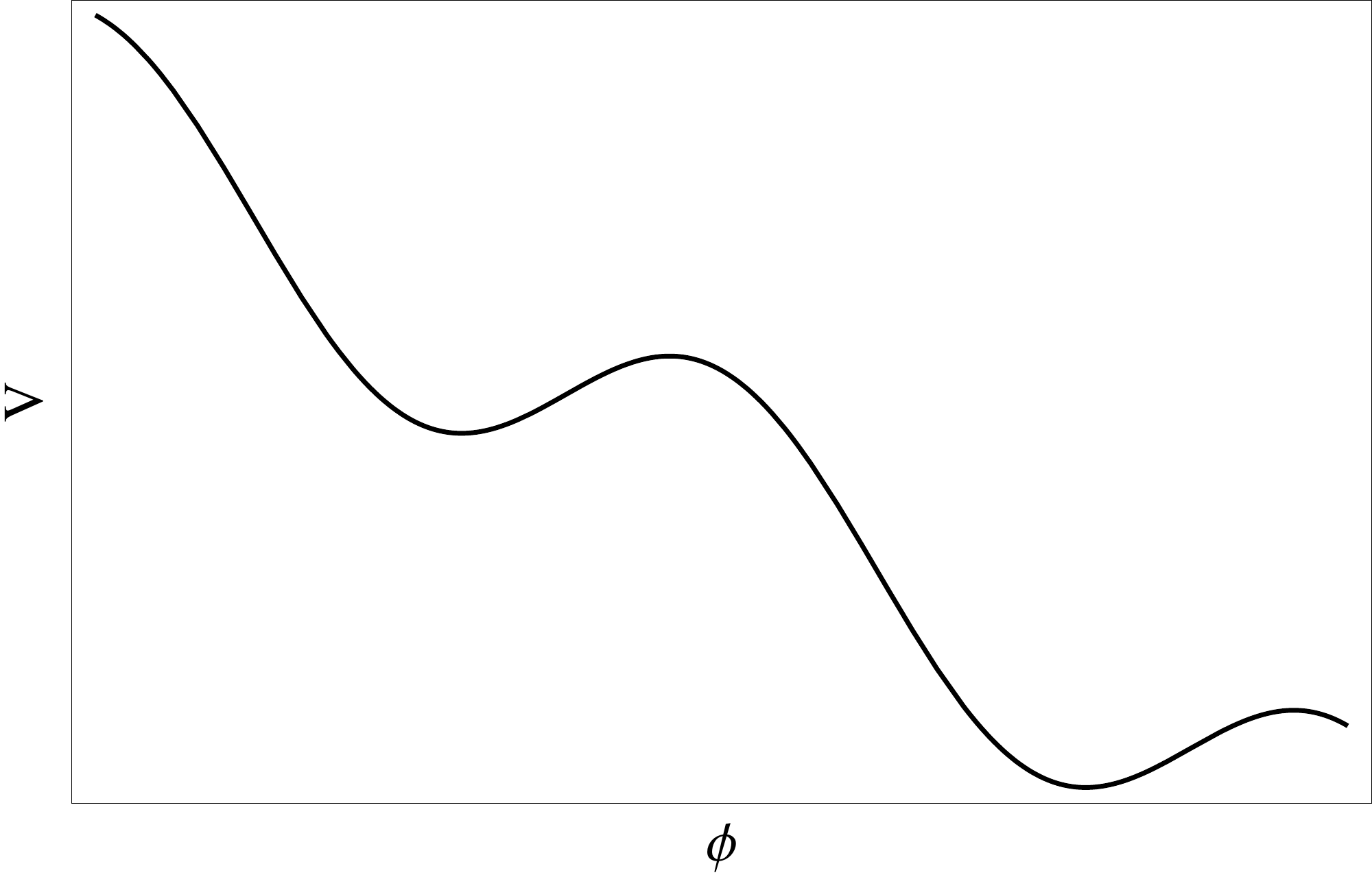}
\caption{A plot of the potential for $\phi$ when Eq.~\ref{Eq: coin} is satisfied.  By eye, one can see that a generic starting point on the potential can result in $\phi$ traveling past all of the subsequent minima.} \label{Fig: potential}
\end{center}
\end{figure}

There is another possibility, where a large $\phi$ initial velocity is not needed in order to scan many Higgs vacua.  As shown in Fig.~\ref{Fig: potential}, if
\bea
\label{Eq: coin}
\epsilon \Lambda^2 f' \sim \Lambda_c^4
\eea
then it is possible that a generic starting point will roll past the first minimum and slowly gather momentum as it rolls past many minima.  Thus a large initial $\dot \phi$ is not needed as long as there is a coincidence of scales shown in Eq.~\ref{Eq: coin}.  If the linear term is too large, then there are no wiggles on top the linear term to stop the $\phi$.  If the linear term is too small, then a generic starting point won't have enough kinetic energy to roll past the first minimum.  Because the friction term scales as $H \sim 1/t$, $\phi$ will roll a non-trivial amount between the time right after inflation to when $\phi$ starts to fast roll past the various minima.  Due to this effect, numerically one finds that one needs these scales to be within $\mathcal{O}(10\%)$ of each other.

Regardless of whether or not $\phi$ started with a small initial pseudo-slow roll velocity or not, $\phi$ is not able to scan its entire field range until $H \sim \epsilon$ simply due to not having enough time.  Since $\phi$ gains $\mathcal{O}(\Lambda^4)$ kinetic energy while scanning its potential, $\dot \phi$ very quickly approaches $\Lambda^2$.   Now the story proceeds as mentioned before.  $\phi$ overshoots all of its minima until $h \sim$ 100 GeV and particle production kicks in.  After losing most of its kinetic energy into particle production, $\phi$ gets stuck in the nearest vacua.  The only difference from before is that $H \sim \epsilon$ rather than $H > \epsilon$ or $H < \epsilon$.  As before, the largest UV cutoff that can be reached with this approach is $\Lambda \lesssim 5 \times 10^4$ GeV.

After $\phi$ finds the correct Higgs vacuum, reheating can occur.  Because $H \sim \epsilon > \Lambda^2/M_p$, most of the energy density in the universe is still in the inflaton.  The only constraint on reheating is that it occurs after $\phi$ finds the correct Higgs vacuum, otherwise the relaxion would have scanned the thermal Higgs mass instead of the zero temperature mass. 

Another possibility is if the inflaton decays into dark sector radiation while the relaxion particle production reheats the visible sector. Because initially the dark sector would have more energy than the visible sector, a period of matter domination would need to occur in the visible sector. This would allow the dark sector radiation to red shift away to acceptably small levels.

As the initial conditions are important, we now detail the constraints on inflation coming from requiring that these initial conditions are reasonable.

\begin{itemize}
\item pseudo slow roll initial conditions
\end{itemize}

If $\phi$ exits inflation while pseudo slow rolling, the decreasing $H$ will only make pseudo-slow rolling easier and it will continue to roll.  The conditions for pseudo-slow rolling during inflation were specified in Sec.~\ref{Sec: during}.  The relevant bound is $H_\text{inflation} \lesssim \frac{\epsilon \Lambda^2}{\Lambda_c^2} $. Taking this approach requires a low scale of inflation.

\begin{itemize}
\item Starting from rest
\end{itemize}

If $\phi$ starts from rest, it should start at a generic point in its potential.  One way for this to occur is if the quantum fluctuations during inflation dominate over the potential energy that attempt to localize $\phi$ at a minimum. This approach requires that 
\bea
H_\text{inflation}^4 \gtrsim V_\phi \sim \Lambda^4
\eea
so that inflation populates all parts of the $\phi$ potential equally~\cite{Linde:1991sk,Starobinsky:1986fx}.  We find that rather than low scale inflation, this approach favors a large scale of inflation.  As mentioned before, in order to avoid setting the Higgs to a random point in its potential, the Higgs should possess the coupling $h h^\dagger R$ so that it obtains a large Hubble sized mass during inflation and is stuck in its minimum.  Additionally, we need $H_\text{inflation} \ll f'$ so that in the entire universe $\phi$ is starting off from roughly the same location, otherwise there are $e^{60}$ different starting locations and some will end up not making it past the first minimum. In this approach $H$ is larger than the cutoff and thus our effective field theory description breaks down. We assume that as long as $H < f'$, one can embed our setup in a inflationary sector with a cutoff larger than Hubble without disrupting the relaxion mechanism (which only operates at lower energies, when $H \sim \epsilon$).

\subsection{Additional concerns}
\label{Sec: additional concerns}

There are several additional issues that could potentially be worrisome.  We briefly detail a few here.

\begin{itemize}
\item Relaxion abundance
\end{itemize}

Like the QCD axion, the relaxion can be produced by both misalignment and thermal scattering.  At temperatures above the weak scale, the relaxion is in thermal equilibrium with the SM plasma through the coupling to weak gauge bosons (e.g. $\phi W \leftrightarrow W Z$ or $\phi Z \leftrightarrow q \bar q$). This process leaves equilibrium for $T < m_W$ when the weak gauge boson number density becomes Boltzmann suppressed.    
If the relaxion is very long lived, it must be lighter than $\sim 10$ eV in order to be consistent with warm dark matter bounds~\cite{Viel:2005qj}.  On the other hand, if the relaxion is heavier than $\sim 10$ MeV, e.g. as is the case in Sec.~\ref{Sec: during}, it can decay sufficiently fast to electrons if one includes the shift symmetric coupling
\bea
\mathcal{L} \supset \frac{\partial_\mu \phi}{f} ( J^\mu_{e^c} - J^\mu_{\tau^c} ),
\eea
where $J^\mu_{e^c}$ ($J^\mu_{\tau^c}$) are the right-handed electron (tau) current.  This combination of currents was chosen in such a way as to not reintroduce a coupling to $F \tilde F$\footnote{To be super safe from astrophysical constraints, one can couple the relaxion to the muon or to another current that does not have an anomaly with the photon.}.  Alternatively, one can simply appeal to a low reheat temperature and not thermally produce the relaxion.

A large relaxion energy density is also produced through the misalignment mechanism.  Once the relaxion loses enough energy to get trapped in one of the minima, it starts oscillating with an $\mathcal{O}(1) \times f'$ amplitude, which leads to a large energy density that redshifts as matter. In the case in which the scanning happens before or during inflation, this energy density is rapidly diluted away by inflation. When the scanning is done after inflation, the energy density in the relaxion field would overclose the universe if
\bea
\frac{\rho_\phi}{s_\gamma} \gtrsim 10^{-10} \text{GeV} \Rightarrow \Lambda_c (\frac{f'}{M_p})^{3/2} > 10^{-10} \, \text{GeV} \ ,
\eea
where $s_\gamma$ is the photon entropy. In this case one needs to make sure that this energy density is either diluted away or transferred to radiation.  Although naively the rate for depleting the condensate through scattering is large for temperatures above the $W$ mass, it has a large suppression due to the approximate cancelation of the bose enhanced scattering of thermal particles into the condensate and the scattering out of the condensate~\cite{Drewes:2013iaa}. This leads to an additional $m_\phi/T$ suppression of the rate for depleting the condensate through scattering.  Assuming that the SM is reheated to temperatures above the weak scale and requiring that the interactions which destroy the condensate occur faster than Hubble at that time, gives the constraint
\bea
\frac{\Lambda_c^2}{f^2 f'} \gtrsim \frac{1}{M_p} \  \Rightarrow \ f' \lesssim \frac{\Lambda_c^2 M_p}{f^2}
\eea

\begin{itemize}
\item Particle production from a changing Higgs vev
\end{itemize}

As seen before, it is extremely important that the initial temperature is small.  As $\phi$ scans the Higgs mass, particles are required not to be produced until Eq.~\ref{Eq: speed} is satisfied.  As the Higgs vev is constantly changing, one might worry that this changing Higgs vev is creating particles.  One can treat this case as simply the case where there is a time dependent mass for the SM fermions.  By dimensional analysis, particles can only be produced when 
\bea
\label{Eq: WKB}
\dot m > m^2
\eea
This statement is equivalent to saying that there is particle production only when the WKB approximation fails.

By massaging Eq.~\ref{Eq: WKB}, one finds that particle production is avoided when
\bea
\label{Eq: y}
\epsilon \dot \phi < y v^3
\eea
where y is some yukawa coupling.  For heavy particles, this equation is always satisfied and heavy particles are never produced.  For particles with small yukawa couplings, Eq.~\ref{Eq: y} isn't always satisfied and particles can be produced.  To estimate the number of particles produced, we consider the process where $m$ becomes small goes through zero and then becomes large again and ask how many particles are produced.  As this process involves changing m even more than what we are doing, this can be considered to be an overestimate of the total number of particles produced.  This question was worked out in Ref.~\cite{Kofman:2004yc}.  They found that the number of particles produced was
\bea
n \lesssim (y \dot m)^{3/2}  \qquad \rho \sim y v n
\eea
Given the constraints on our cutoff scale, we find that the final temperature associated with the produced particles satisfies $T \ll v$, so that the relaxion dynamics is not changed significantly by particles produced due to the changing Higgs vev.  Thermal effects on the Higgs mass are also small.

\begin{itemize}
\item Erasure of $\Lambda_c$ cosine potential
\end{itemize}

Another thermal effect one might be worried about is what happens if $\phi$ is reheated to a temperature where the bumps $\Lambda_c$ disappear.  To be more precise, if the sector whose dynamics is responsible for the bumps is also reheated to a temperature larger than $\Lambda_c$ then the bumps will disappear.  Assuming that the bumps are generated by $(\phi/f') G_h \tilde G_h$, where $G_h$ is the field strength of a hidden confining sector, and that the SM temperature is $\sim \sqrt{\dot \phi}$, one finds that the hidden sector will not equilibrate with the SM as long as
\bea
\label{Eq: hidden-eq}
\frac{\dot \phi^{5/2}}{f'^2 f^2} \lesssim H.
\eea
An additional constraint comes from requiring that the energy dumped into the $G_h \tilde G_h$ sector is smaller than $\Lambda_c^4$.  These constraints are satisfied in all the scenarios we discuss.

\begin{itemize}
\item Higgs tracking its minimum as Higgs mass is being scanned
\end{itemize}

Another important effect is that the Higgs efficiently tracks the minimum of its potential as its mass is being scanned.  We can study how well tracking happens by expanding around the tracking solution.  
The equations of motion for the Higgs field is
\bea
\ddot h = (\Lambda^2 - \epsilon \phi) h - \lambda h^3 = (\Lambda^2 - \epsilon \dot \phi t ) h - \lambda^3 h
\eea
Using the field redefinitions
\bea
h' = \frac{\sqrt{\lambda}}{\Lambda} h \qquad \epsilon' = \epsilon \frac{\dot \phi}{\Lambda^3} \qquad \tau = \epsilon' \Lambda t
\eea
and expanding around the solution
\bea
h'(\tau) = \sqrt{1-\tau} (1 + \delta(\tau) ) ,
\eea
the new equations of motion are
\bea
\label{Eq: bessel}
\frac{1 + \delta}{9} = x  \frac{d \delta}{dx} + x^2  \frac{d^2 \delta}{dx^2} + x^2 \delta (1 + \delta ) ( 1+ \frac{\delta}{2})  \qquad x = \frac{2^{3/2}}{3 \epsilon'} (1-\tau)^{3/2}
\eea
We are interested in the limit of small $\delta$ so that the Higgs tracks its minimum well.  In this limit, Eq.~\ref{Eq: bessel} becomes the nonhomogeneous Bessel's equation.  In the large $x$ limit, the solutions oscillate with an amplitude that goes as $1/x^{1/2}$.  From this we see that the large $x$ limit automatically leads to a small $\delta$.

Requiring that the expansion be good, $x \gg 1$, gives
\bea
(1 - \tau)^3 \gtrsim \epsilon'^2
\eea
The Higgs stops tracking the minimum when the perturbation theory breaks down.  This happens around
\bea
h = \frac{1}{\sqrt{\lambda}} \Lambda h' \sim \Lambda \epsilon'^{1/3} \sim ( \epsilon \dot \phi )^{1/3} < v
\eea
where we have now required that this failure of tracking happens at Higgs vevs below the 100 GeV minimum.  This constraint is satisfied for all of the models considered.

\begin{itemize}
\item Thermalization of the produced particles
\end{itemize}

When treating the system after particles have been produced, we have taken the finite density system to be thermal.  In particular, we assume that the system is thermal on time scales of order the inverse tachyon mass.  To estimate at what energy densities the system is first well described by a thermal ensemble, we compare the rate of thermalization with the tachyon time scale.  Initially the produced particles have energy of order their mass ($v$) and the time scale of the tachyon is of order the mass of the W/Z bosons.  It is appropriate to treat the system as thermal when
\bea
\Gamma(W^+ W^- \rightarrow e^- e^+) \sim n \langle \sigma \mathrm{v} \rangle\sim \frac{\rho_\text{W/Z}}{v} \frac{1}{v^2} > m_\text{tachyon} \sim v
\eea 
We see that once the relaxion has dumped more than $v^4$ energy density into the EW gauge bosons, the system thermalizes quickly and can be accurately treated as a thermal system.  As the total energy dumped into the system is $\Lambda^4 \gg v^4$, we see that the system very quickly reaches the point where it can be accurately treated as thermal.

\begin{itemize}
\item Coupling with the photon
\end{itemize}

As shown in App.~\ref{App: LR}, a lack of a coupling with the photon can naturally arise in various UV models.  However, couplings to the photon can still be generated through shift symmetry breaking couplings suppressed by $\epsilon$.  We now briefly estimate exactly how small the coupling with the photon needs to be in order not to ruin the mechanism.  The main problem with producing photons is that if $\rho_\gamma > v^4$, finite density/temperature effects will make it so that the Higgs mass being scanned is not the vacuum mass and the wrong Higgs mass will be selected.

One conservative constraint that applies to all cases is to require that the time scale for the production of photons takes longer than Hubble.  After a Hubble time, the Higgs will have either have found its minimum, or the energy density in photons will have diluted away.
\bea
\mathcal{L} \supset \frac{\phi}{f_\gamma} F_\gamma \tilde F_\gamma \qquad \frac{\dot \phi}{f_\gamma} \sim \frac{v f}{f_\gamma} < H
\eea
When maximizing the UV cutoff, one typically saturates Eq.~\ref{Eq: H}.  When saturated, we find that $ f_\gamma > f \frac{M_p v}{\Lambda^2} \gtrsim 10^{10}  f$.  As shown in App.~\ref{App: LR}, a PQ symmetry can forbid this coupling.  However, $\epsilon$ interactions are a soft breaking of the PQ symmetry so that we cannot forbid operators such as
\bea
\mathcal{L} \supset \frac{\epsilon \phi}{f_\gamma \Lambda} F_\gamma \tilde F_\gamma
\eea
Using the fact that field excursions are sub-Planckian, Eq.~\ref{Eq: epsilon}, we find that as long as the UV cutoff is above the weak scale, then $\epsilon$ suppressed production of photons is negligible.

\section{Variations of Relaxation}
\label{Sec: variants}

In this section, we discuss some minor variants of the previously discussed model of relaxation.  The point of these variants is to explore slightly different mechanisms for implementing relaxation as well as showing how small variations in the model can lead to conceptually different interpretations of the same model.  We first discuss a variant for relaxation before inflation and then discuss a variant for relaxation after inflation.

\begin{itemize}
\item Before inflation
\end{itemize}

When implementing relaxation before inflation, there were a set of strange initial conditions.  We can also start with a more ``natural" set of initial conditions in which both the Higgs and $\phi$ are displaced from their minima and with a non-trivial thermal background. The only problem in this case is that one would naturally expect that $\rho_\text{thermal} \sim \Lambda^4$.  Generically this means that the Higgs mass which the relaxion responds to is the thermal mass and not the vacuum mass.  This difference is important if the temperature of the bath is above 100 GeV. The simplest way around this problem is to assume that at energies above the electroweak scale there is a very large number of new degrees of freedom:
\bea
g_\star \gtrsim \frac{\Lambda^4}{v^4}
\eea
If this is the case, then the thermal mass is not appreciably different than the vacuum mass.  Since the relaxation time of the Higgs to its minimum is quick ($t \sim 1/\Lambda$), the Higgs relaxes to its minimum much faster than $\phi$ scans the Higgs mass. In this case one does not have to worry about erasing the cosine potential for $\phi$ since the temperature is much smaller than the barrier height $\Lambda_c$.
As a simple example of this scenario, consider the case where there are simply a large number of axion like particles coupled to the standard model.

The presence of a large $g_\star$ forces us to change the stopping mechanism slightly.  As emphasized before, thermal effects were critical in stopping the $\phi$ field.  Having a large $g_\star$ forces temperature based effects to be too small to be of interest.  In this case what happens is that exponential particle production forces the velocity of $\phi$ to track
\bea
\dot \phi \sim f m_A
\eea
$\phi$ continues to fly over the wiggles until $\dot \phi \sim \Lambda_c^2$.  At this point, the potential energy dominates over the kinetic energy and $\phi$ becomes stuck in a minimum.  Thus we find that this approach to relaxation forces the new constraint that
\bea
f v \sim \Lambda_c^2
\eea
Depending on the value of $g_\star$, the UV cutoff can vary between $5 \times 10^4$ GeV and $10^6$ GeV. This requires an exponentially large number of new degrees of freedom near the weak scale (which could all be part of a particle rich hidden sector).

\begin{itemize}
\item After inflation
\end{itemize}

One of the most intriguing aspects of relaxation after inflation is the connection between the reheat temperature and obtaining the correct Higgs mass.  If the initial conditions have $m_H^2 < - 125^2$ GeV, then the story proceeds as described before.  $\mathcal{O}(1)$ of the relaxion potential energy is used to reheat the SM to temperatures of order the UV cutoff $T_\text{high} \sim \Lambda$.  If the initial conditions have $m_H^2 > - 125^2$ GeV, then the universe is reheated to a much lower temperature as particle production occurs before the relaxion can obtain a large kinetic energy.  We will call this lower reheat temperature $T_\text{low}$.

One can take advantage of these differing reheat temperatures to obtain a variant of relaxation which has conceptually different interpretations from the original models.  Models involving freeze-out and decay baryogenesis and models of thermal dark matter all have a distinct temperature where freeze out occurs ($T_\text{DM,Baryo} \sim m/20$).  If one reheats to a temperature below this temperature, then the particle is not produced.  The correct dark matter abundance or baryogenesis is thus not obtained.  If one reheats to a temperature above this temperature, then baryogenesis and thermal dark matter proceeds as usual.  

Now imagine that $T_\text{high} > T_\text{DM,Baryo} > T_\text{low}$.  Thus if one starts with the correct initial conditions to get the correct Higgs vev, one also has baryogenesis and the correct dark matter abundance.  If one starts off with the ``wrong" initial conditions, then one gets the incorrect Higgs vev and also does not have baryons nor dark matter.  This model has several different interesting interpretations.

The first interpretation of the model is the relaxion interpretation.  A large set of initial conditions all flow towards the correct Higgs vev as per the usual relaxion approach.  Unlike the original relaxion models, regions of space where the incorrect initial conditions were present do not have matter.

The second interpretation of our model is anthropic.  In order for the anthropic argument for the cosmological constant to go through, the universe must have matter in it and thus baryogenesis and dark matter genesis must have occurred.  Because we have connected reheating, baryogenesis and dark matter with obtaining the correct Higgs vev, anthropics applied to the cosmological constant also gives us the correct Higgs mass as a consequence.  

A third and final interpretation of our model is one we call the ``conditional" interpretation.  We have constructed a model where $P( \langle h \rangle  \sim 256 \, \text{GeV} \, | \, \text{baryogenesis and dark matter} ) \sim 1$.  Since we have observed a universe with baryons and dark matter, the Higgs mass must be 125 GeV.  This approach is similar in spirit to anthropics but with a critical difference.  Anthropics is conditional on the existence of observers.  Calculations supporting anthropics can not be done convincingly as we have no theory for the measure of the multiverse and we have no real understand of when life can develop.  In contrast, our conditional approach is entirely calculable.  Reheating is a QFT problem that is well understood and can be studied in the $M_p \rightarrow \infty$ limit.  Anthropics is simply the conditional approach applied to something we do not understand so that the only difference between these two approaches is calculability.

\section{Discussion}
\label{Sec: discussion}

In this paper, we have used particle production to implement the relaxion approach.  We have shown that relaxation using particle production can be done before, during and after inflation.  The fact that relaxation can be done after inflation is extremely exciting because this means that inflation can be completely decoupled from the story.  In fact, implementing relaxation after inflation prefers large scale inflation.
We also presented an interesting implementation of relaxation after inflation which had several different interesting interpretations, distinct from the usual relaxion solution to the Hierarchy problem.

One might try to implement relaxation with production of fermions.  In this case it is much more difficult because the Pauli exclusion principle makes exponential production of fermions difficult.  Scattering effects are critical in order to deplete the fermion number density so that fermion production can occur continuously.

If one is not careful, particle production can also happen in other relaxion models.  This is because coupling the relaxion to a QCD-like sector will always also couple the relaxion to the photon due to mixing effects with the mesons.  Exponential particle production can occur and if the steady state temperature is too large, the QCD-like sector may deconfine or the Higgs mass being scanned might be the thermal mass instead of the vacuum mass.  It is simple to see that this constraint is strongest when one has sub-Planckian field excursions.

In summary, relaxation is still an approach to the hierarchy problem which has not been fully explored.  We hope that future investigation will find even more and exciting ways to solve the hierarchy problem.

\acknowledgments
We thank Nikita Blinov, Peter Graham, Shamit Kachru, David E Kaplan, Jeremy Mardon, Lorenzo Ubaldi and Lian Tao Wang for discussion.  
A.H. and G.M.T. are supported by the DOE Grant DE-SC0012012 and NSF Grant 1316699. We thank the Aspen Center for Physics, which is supported by National Science Foundation grant PHY-1066293, where part of this work was completed.

\appendix

\section{Bound for slow-roll models}
\label{App: N}

In this appendix we derive a bound on the minimal number of e-foldings required to effectively scan the Higgs mass in general relaxion models in which the field scanning the Higgs mass is slow-rolling down an approximately linear slope. This bounds applies to models in which the relaxion stops at a point in which the slope of a cosine barrier becomes comparable to the approximately linear slope associated with the shift symmetry breaking potential. In the slow-roll regime we have
\be
\dot \phi = - \frac{V'(\phi)}{3 H} \approx - \frac{\epsilon \Lambda^2}{3 H}
\ee
Therefore the number of e-foldings required in order to have a field excursion of size $\Lambda^2 / \epsilon$ is given by
\be
\Delta N_e \approx  \frac{ \Lambda^2 / \epsilon}{\dot \phi / H} \approx \frac{H^2}{\epsilon^2}
\ee

The conditions for exiting the pseudo-slow roll were given in Sec.~\ref{Sec: during}.
\bea
\dot \phi \approx \frac{\epsilon \Lambda^2}{H} \lesssim \Lambda_c^2
\eea
where $\Lambda_c^4$ is the amplitude for the cosine potential of $\phi$.  Using this, we find that the number of e-foldings is bounded by
\bea
\Delta N_e \gtrsim \frac{\Lambda^4}{\Lambda_c^4}
\eea
If $\Lambda_c < v$ as in the original relaxion scenario, one automatically requires a very large number of e-foldings of inflation. This large number of e-foldings can be reduced by taking into account the effect of non-constant Hubble~\cite{Patil:2015oxa}.

\section{A specific left right symmetric model}
\label{App: LR}

In this appendix we very briefly write down a more explicit way a left right symmetric model can naturally avoid giving the relaxion a coupling to the photon.  Consider the minimal left right symmetric model where we add the $SU(2)_R$ gauge bosons and the right handed Higgs field.  To this, we add vector-like doublets $\psi_L$ (charged under $SU(2)_L$) and $\psi_R$ (charged under $SU(2)_R$).  The PQ charges of these two fields are equal and opposite as required by parity.  Additionally, we impose that the CP odd and P even relaxion has a shift symmetry under this weak PQ symmetry.  The Lagrangian for these new fields is
\bea
\mathcal{L} \supset \Lambda e^{i \phi/f} \psi_L \psi_L^c + \Lambda e^{-i \phi/f} \psi_R \psi_R^c \qquad \Rightarrow \qquad \mathcal{L} \supset - \frac{1}{4 g_L^2} W_L^2 - \frac{1}{4 g_R^2} W_R^2 + \frac{\phi}{f} (W_L \tilde W_L - W_R \tilde W_R)
\eea
In order to demonstrate that this coupling is robust against finite threshold corrections and all other sorts of perturbative phenomenon, we will utilize the PQ symmetry
\bea
\phi \rightarrow \phi + \alpha \qquad \theta_L \rightarrow \theta_L - \alpha \qquad \theta_R \rightarrow \theta_R + \alpha
\eea
The IR couplings of the relaxion can be entirely determined from the how $\theta_{L,R}$ appear in the IR.  $\phi$ appears in such a way that the couplings are invariant under the PQ symmetry.  The spurions $\theta_{L,R}$ are special because they couple to total derivatives and are topological parameters.  This means that there are no perturbative phenomenon which can change how they appear in the Lagrangian.  To get the coupling to the photon, one simply need to express the $\theta_L W_L \tilde W_L + \theta_R W_R \tilde W_R$ in terms of the photon.  Note that because the gauge couplings are in front of the kinetic terms, the overlap between $W_L$ and $W_R$ with the photon is completely determined by the requirement that the vevs of the left handed and right handed Higgses are not charged.  In this basis, the overlap is RG invariant as RG flow cannot change the zero electric charge of the Higgs vevs.  Because the left and right handed Higgses have the charges related by parity, we find that in the IR that $\theta_{L,R}$ couples to the photon as
\bea
\mathcal{L} \propto (\theta_L + \theta_R) \gamma \tilde \gamma
\eea
This combination is invariant under the PQ symmetry so that $\phi$ cannot couple to the photon.
As the structure of this coupling comes from the PQ symmetry, as long as there are no operators which break this PQ symmetry, $\phi$ will avoid any couplings with the photon.  In fact, as long as this left right symmetric model is put right near the cutoff, the requirement that the relaxion does not couple to the photon is simply the result of a symmetry and requiring that the UV completion respect the symmetry insures the lack of a coupling to the photon.  The soft PQ and parity breaking parameter $\epsilon$ is small enough so that any coupling with the photon suppressed by $\epsilon$ is unimportant.

\bibliography{ref}

\end{document}